\documentclass[twocolumn,preprintnumbers,amsmath,amssymb]{revtex4}

\usepackage{graphicx}% Include figure files
\usepackage{dcolumn}% Align table columns on decimal point
\usepackage{bm}% bold math
\usepackage{color}% bold math
\begin{document}

\preprint{APS/123-QED}

\title{Magnetothermoelectric properties of Bi$_2$Se$_3$}
\author{Beno\^it Fauqu\'e$^{1,\footnote{benoit.fauque@espci.fr}}$, Nicholas P.  Butch $^{2,\footnote{Present address :  Condensed Matter and Materials Division Lawrence Livermore National Laboratory, Livermore, CA 94550 USA }}$, Paul Syers$^{2}$,  Johnpierre Paglione $^{2}$, Steffen Wiedmann$^{3}$, Aur\'elie Collaudin$^{1}$, Benjamin Grena$^{1}$, Uli Zeitler$^{3}$, }

\author{Kamran Behnia$^1$}%

\affiliation{
$^1$ LPEM (UPMC-CNRS), Ecole Superieure de Physique et de Chimie Industrielles, 75005 Paris, France\\
$^2$ Center for Nanophysics and Advanced Materials, Department of Physics, University of Maryland, College Park, Maryland 20742, USA\\
$^3$ High Field Magnet Laboratory, Institute for Molecules and Materials, Radboud University Nijmegen, Toernooiveld 7, 6525 ED Nijmegen, The Netherlands\\
}


\begin{abstract}
We present a study of entropy transport in Bi$_2$Se$_3$ at low temperatures and high magnetic fields.  In the zero-temperature limit, the magnitude of the Seebeck coefficient quantitatively tracks the Fermi temperature of the 3D Fermi surface at $\Gamma$-point as the carrier concentration changes by two orders of magnitude (10$^{17}$ to 10$^{19}$cm$^{-3}$). In high magnetic fields, the Nernst response displays giant quantum oscillations indicating that this feature is not exclusive to compensated semi-metals.  \color{black}A comprehensive analysis of the Landau Level spectrum firmly establishes a large $g$-factor in this material and a substantial decrease of the Fermi energy with increasing magnetic field across the quantum limit. Thus, the presence of bulk carriers significantly affects the spectrum of the intensively debated surface states in Bi$_2$Se$_3$ and related materials.\color{black}
\end{abstract}
\pacs{Valid PACS appear here}% PACS, the Physics and Astronomy
\maketitle

The Bi$_2$X$_3$ family (X= Se, Te) is attracting tremendous attention as a Topological Insulator (TI). Recently, the existence of this class of bulk insulators was predicted and confirmed\cite{Hasan}. In a TI, the bulk energy gap is traversed by spin polarized surface states. Therefore, the electrical conduction is expected to occur only at the surface. In practice, however, these materials are often low-density bulk metals. Interestingly, many of the TI of the first and the second generation are well-known thermoelectric materials\cite{NolasBook} and present a sizeable thermoelectric figure of merit. This quantity,  \color{black} $ZT=\frac{S^2T}{\kappa\rho}$ \color{black}(here $S$ is the Seebeck coefficient, $\kappa$ is thermal conductivity and $\rho$ is resistivity) characterizes the thermoelectric efficiency of a material. To this date, the largest  thermoelectric figure of merit in a bulk material at room temperature has been reported in Bi$_2$Te$_3$ (ZT=0.8 at T=300K \cite{Goldsmith, Mahan}).

In spite of the fundamental and technological interest in Bi$_2$Se$_3$ and Bi$_2$Te$_3$, their thermoelectric response has not been investigated at temperatures low enough to distinguish between the two competing (semiconducting vs. metallic) ground states. In this paper, we report on measurements of Seebeck and Nernst effect  in n-type Bi$_2$Se$_3$  with a bulk carrier concentration varying from 10$^{19}$cm$^{-3}$ to 10$^{17}$cm$^{-3}$ at low temperature and in a magnetic field  as strong as 32 T. We find that the low-temperature magnitude of the Seebeck coefficient is set by the Fermi temperature, which can be directly extracted from the measured properties of the metallic Fermi surface. The transverse thermoelectric (Nernst) response displays large quantum oscillations. Such oscillations were previously reported in low-density compensated semi-metals, bismuth and graphite. Their observation in a single-band uncompensated system indicates that they are generic to any low-carrier system pushed to the quantum limit by a sufficiently strong magnetic field. Analysis of quantum-oscillations allows us to: i) document a continuous field-induced shift in chemical potential affecting the periodicity of the oscillations; ii) quantify the magnitude of Zeeman splitting in this system, which is expected to host a large spin-orbit coupling. \color{black} These two points are crucial in discussing the Landau spectrum of bulk states. They also affect the analysis of the Landau spectrum of the surface carriers, which share the same chemical potential with bulk.
\color{black}

\subsection{ Quantum oscillations in Seebeck and Nernst effect }
 \begin{figure}[htbp]
\begin{center}
\includegraphics[angle=0,width=9.0cm]{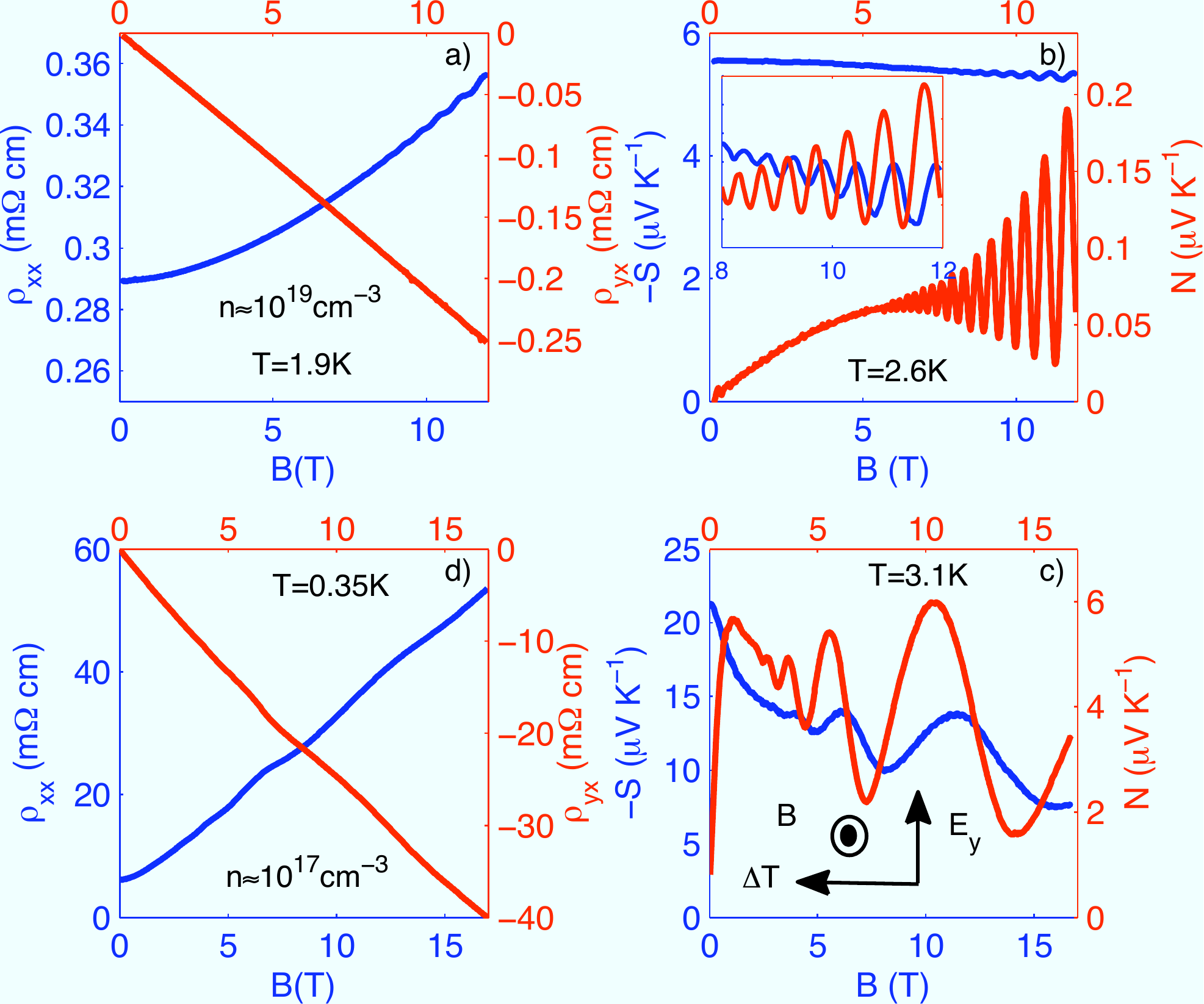}
\caption{ (Color online) Electrical and entropy transport measurements of Bi$_2$Se$_3$ for the samples A$_1$ ($n(A_{1})$ $\approx $1e19 $cm^{-3})$ and B$_{1}$ ($n(B_{1})$ $\approx$ 1e17 $cm^{-3}$)  a) $\rho_{xx}$ (in blue) and $\rho_{yx}$ (in red) of A$_1$ as a function of the magnetic field for T=1.9 K b) $S$ (in blue) and $N$ (in red) of A$_1$ as a function of the magnetic field  for T=2.6 K (the insert show a zoom of  $S$ and $N$ between 8 and 12T) c) $\rho_{xx}$ (in blue) and $\rho_{yx}$ (in red) of B$_1$ as a function of the magnetic field for  T=350mK d) $S$ (in blue) and $N$ (in red) of B$_1$ as a function of the magnetic field  for  T=3.1K }
\label{Fig1}
\end{center}
\end{figure}

Measurements of longitudinal (S=$\frac{E_x}{\Delta_x T}$) and the transverse (N=$\frac{E_y}{\mid \Delta_x T \mid}$) thermoelectric response  were performed on a standard one-heater-two-thermometers setup. For all samples, the electric current and the thermal gradient were applied in the plane  of the quintuple layers  and the magnetic field was oriented along the trigonal direction (perpendicular to the layers). The samples used in this study are similar to those previously studied by  longitudinal and transverse magnetoresistance experiments \cite{Butch} and described there.

In Fig.1, we compare the electrical transport and the entropy transport in two samples A$_1$ and B$_1$ with typical bulk concentrations of  10$^{19}$cm$^{-3}$ to 10$^{17}$cm$^{-3}$.  For both samples, $\rho_{xx}$ and $\rho_{yx}$ display  quantum oscillations on top of an almost linear monotonic base as previously reported and discussed \cite{Butch,Kohler,Ando,Analtyis}. In presence of a field of about 10 T, we find  for the two concentrations that $\rho_{xx} \approx \rho_{yx}$ whereas  S $>>$ N (the typical ratios are 100 for A$_1$ and 10 for B$_1$). As in the case of the electrical transport, S and N  display quantum oscillations with a phase shift of $\frac{\pi}{2}$. The oscillations are particularly pronounced in N (in the case of A$_1$ they dominate the signal). The surprising sensitivity of the Nernst effect to quantum oscillations in Bi$_2$Se$_3$ is indeed reminiscent of those reported in bismuth \cite{behnia2} and graphite \cite{zhu1}. The origin of these giant quantum oscillations is the subject of on-going theoretical researches. \cite{Bergman2010,Sharlai2011,Lukyanchuc2011}. These two elemental semi-metals are, however, compensated systems, whereas the (bulk) Fermi surface of  Bi$_2$Se$_3$, as we will see below is a single band at the $\Gamma$-point. Therefore, the observation reported here establishes that this effect is not exclusive to compensated systems. We note that there are several reports on Nernst quantum oscillations with a large amplitude and a small frequency in metals like zinc \cite{{Bergeron59}} and aluminum \cite{Fletcher83} or in doped semiconductors such as n-type InAs\cite{Gadzhialiev69} and iron-doped HgSe \cite{Tieke96}.
\begin{figure}
\resizebox{!}{0.42\textwidth}{\includegraphics{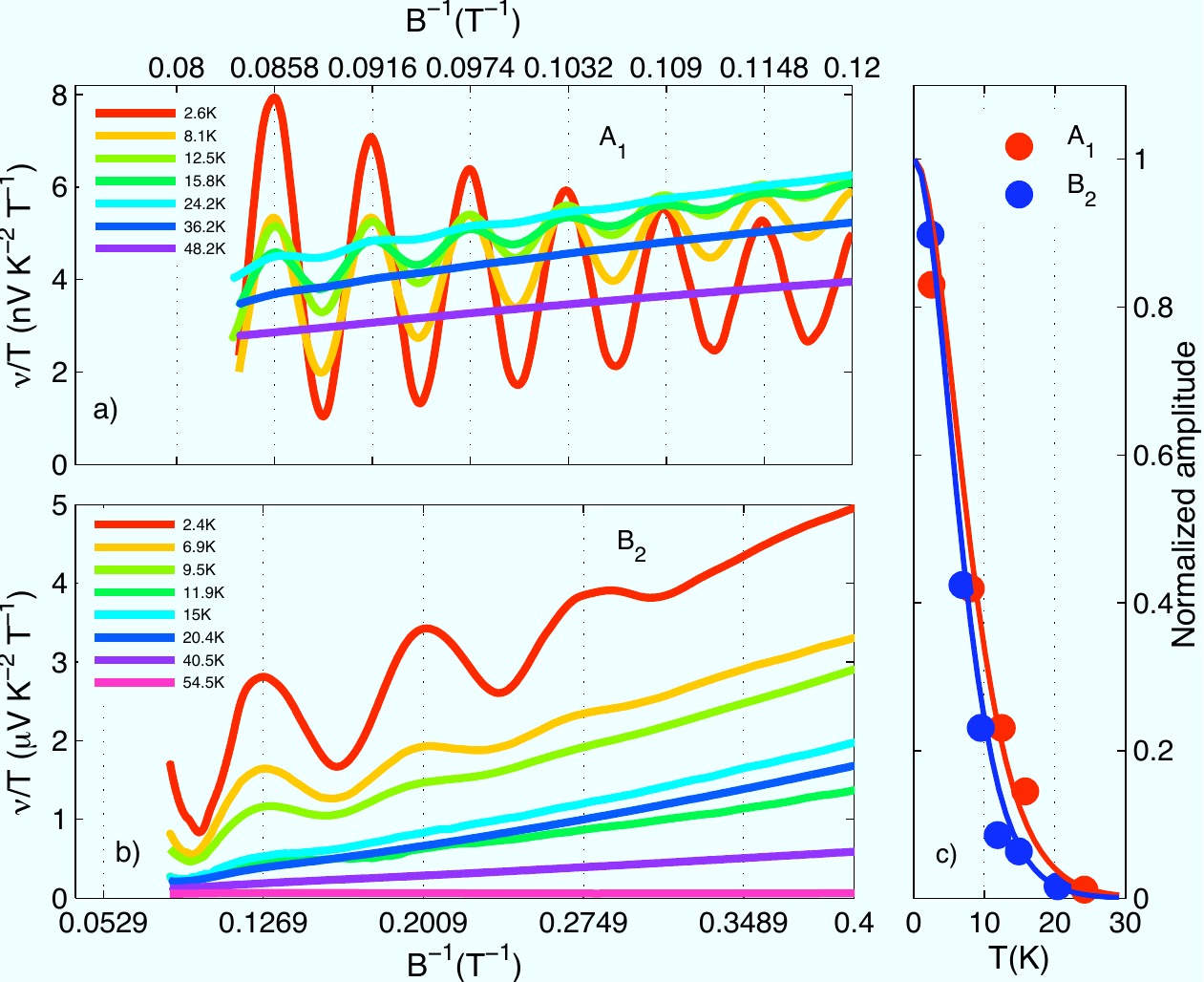}}\caption{ (Color online) a) and b)  $\frac{\nu}{T}=\frac{N}{TB}$ as a function of B$^{-1}$ for the sample A$_1$ and B$_2$ (sample different of B$_1$ with a similar bulk concentration) for the various temperatures explored c) Temperature dependence of the amplitude of the oscillations of the sample A$_1$ (in red) and B$_2$ (in blue). The line correspond to a fit using the Lifshitz-Kosevich formula (see the text ) }
\label{Fig2}
\end{figure}

\subsubsection{ Effective masse and Dingle Temperature  }
Next, we focus our attention on the temperature dependence of the quantum oscillations in the thermoelectric properties, which have been so far poorly studied.  In Fig.\ref{Fig2}, $\frac{\nu}{T}=\frac{N}{TB}$ is plotted as a function of B$^{-1}$ for various temperature. In both cases, the signal is periodic in B$^{-1}$. The difference of the period between the two samples simply reflects the difference in the carrier concentrations. In both cases, the oscillations disappear at a typical temperature of 20 K suggesting rather similar low effective mass for both concentrations.  Following the standard Lifshitz-Kosevich theory \cite{Shoe}, we fit the temperature dependence of the oscillating part of $\frac{\nu}{T}$ to :  $RT = \frac{X}{sinh(X)}$ where $X$ =$\frac{14.69m^*T}{B}$. The data and the fit are shown in Fig.\ref{Fig2}.c).  We found respectively $m^*(A_1)/m_0=0.20\pm0.03$ and $m^*(B_2)/m_0=0.18\pm0.03$ in good agreement with the cyclotron mass deduced by Shubnikov-de Haas oscillations \cite{Butch,Ando,Analtyis}.

% Dingle temperature
  \begin{figure}[h]
\resizebox{!}{0.30\textwidth}{\includegraphics{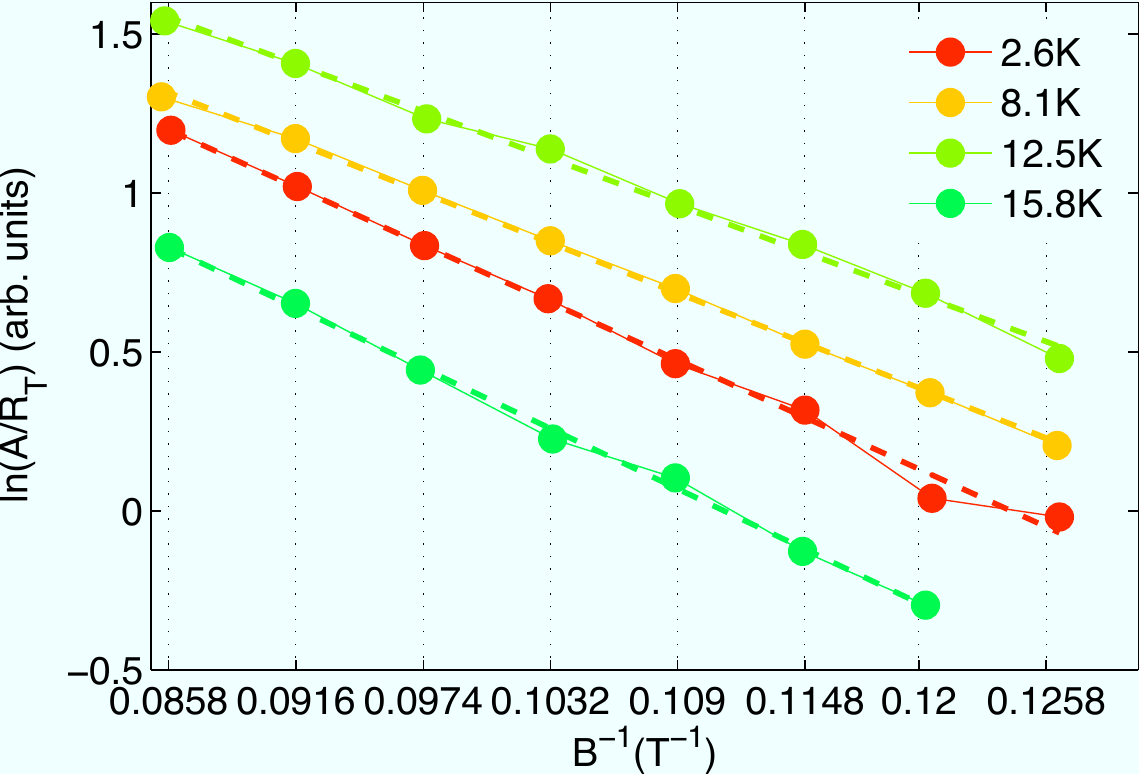}}\caption{  (Color online) "Dingle plot" :  $\ln(\frac{A}{R_T})$ as a function of B$^{-1}$. A correspond to the amplitude of the oscillations as observed in $\frac{\nu}{T}$ and $RT = \frac{X}{sinh(X)}$ where $X$ =$\frac{14.69m^*T}{B}$ (m$^*$=0.2m$_0$ in the case of A$_1$) }
\label{FigDingleSI}
\end{figure}

By studying the field dependence, we can also extract additional information such as a Dingle temperature. In Fig.\ref{FigDingleSI}, we report the so-called "Dingle plot" for four temperatures T=2.8 K, 8.3 K, 12.5 K and 15.8 K for the sample A$_1$. In the context of the Lifshitz-Kosevich formalism \cite{Shoe} , we expect that :  
\begin{equation} \ln(\frac{A}{R_T})=-\alpha T_D m^*\times \frac{1}{B}
\label{TD}
\end{equation}

For the four temperatures, the field dependence of $\ln(\frac{A}{R_T})$ is linear. From Eq.\ref{TD}, we found a Dingle temperature T$_D$=$10\pm1K$ in good agreement with Shubnikov-de Haas measurements \cite{Butch,Kohler,Ando,Analtyis}.  \color{black} The remarkable simplicity of our analysis of the quantum oscillations $\frac{\nu}{T}$ is related to the absence of a phonon dragg contribution in the transverse thermoelectric response (contrary to its significant contribution in the Seebeck effect). This point is consistent with the early analysis of B.Tieke et al. [17] in the case of iron-doped HgSe.
\color{black}

% Amplitude du signal: 
\subsubsection{ Magnitude of the Seebeck and Nernst effect  }

\begin{figure}
\resizebox{!}{0.66\textwidth}{\includegraphics{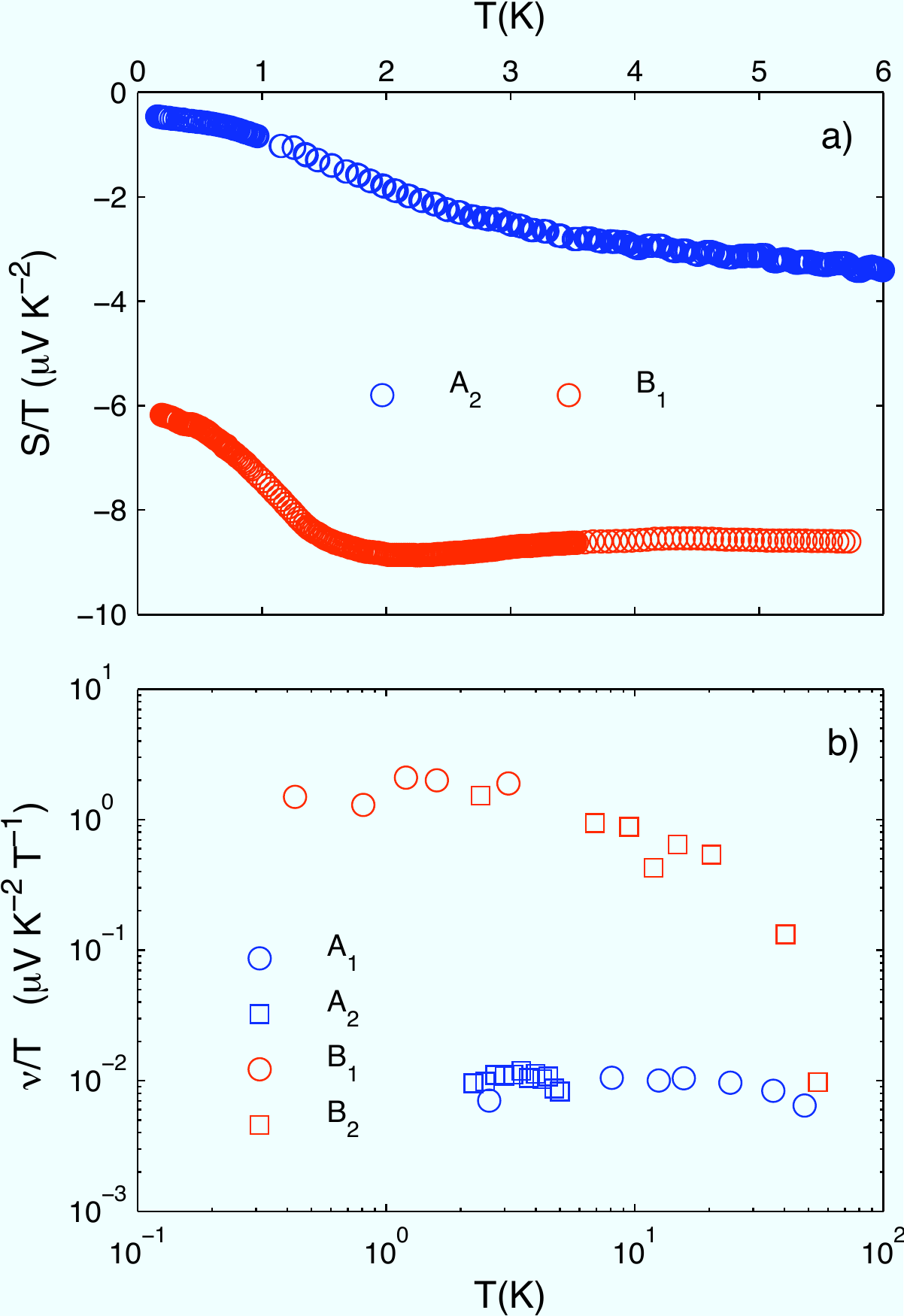}}\caption{ (Color online) a)  $\frac{S}{T}$ as a function of the temperature for A$_2$ (same batch as A$_1$ with a slightly lower concentration) and B$_1$ b) $\frac{\nu}{T}$ at B=1T  as a function of the temperature for A$_1$, A$_2$,B$_1$ and B$_2$}
\label{Fig3}
\end{figure}

The magnitude of the thermoelectric response is dramatically affected by the change in the carrier concentration.  Fig.\ref{Fig3} a) and b),  present  $\frac{S}{T}$ and  $\frac{\nu}{T}$ as a function of temperature for the four samples studied. As the doping passes from 10$^{19}$ cm$^{-3}$ to 10$^{17}$ cm$^{-3}$, $\frac{S}{T}$ increases by one order of magnitude and  $\frac{\nu}{T}$ increases by two orders of magnitude. In the Fermi liquid picture for a one-band system,  the diffusive Seebeck is expected to be T-linear in the zero-temperature limit with a magnitude proportional to 1/T$_F$ :
 \begin{equation}
\frac{S}{T}=-\frac{\pi^2}{2}\frac{k_B}{e}\frac{1}{T_F}
\label{Eq1}
\end{equation}

Similarly the solution of the Boltzmann equation for the Nernst response  leads to : $\frac{\nu}{T}$=-$\frac{\pi^2}{3}\frac{k_B^2T}{m^*}\frac{\partial \tau}{\partial \epsilon}\mid_{\epsilon=\epsilon_F}$ which can be simplified and rewritten as \cite{BehniaNernst} : 

 \begin{equation}
\frac{\nu}{T}\approx283\frac{\mu}{T_F}[\mu V .K^{-2}. T^{-1}]
\label{Eq2}
\end{equation}

Empirically, these simple equations give a rough account of the magnitude of transport coefficients across several orders of magnitude \cite{BehniaNernst,BehniaSeebeck}.

Bi$_2$Se$_3$ provides a particularly compelling opportunity to check the robustness of Eq. \ref{Eq1} and  Eq. \ref{Eq2} in a wide window of carrier concentration with a barely changing Fermi surface topology. The mobility and the Fermi energy can both be extracted from the data on quantum oscillations. Tab.\ref{TabI} lists  the values of  $\frac{S}{T}$ and  $\frac{\nu}{T}$ obtained at the lowest temperature measured ( 2K for A$_1$ and B$_2$ and 150 mK in the case of  A$_2$ and  B$_{1}$) respectively.  Values for the Fermi temperature and the mobility extracted from experiment were plugged in Eq.\ref{Eq1} and Eq.\ref{Eq2}. The expected values for the four samples are reported in the two last lines of  Tab.\ref{TabI}. The overall agreement between the measurement and the expected values is remarkable. This is particularly the case for the two samples studied at the lowest temperature (A$_2$ and B$_{1}$).

 \begin{table}
 \caption{\label{tab:table1} Properties of Bi$_2$Se$_3$}
 \begin{ruledtabular}
\begin{tabular}{lclcccccl}
Sample &A$_1$\footnote{measured down to 2K}&A$_2$\footnote{measured down to 0.15K}&B$_{1}$$^b$&B$_2$$^a$\\
\hline
$\mu$(cm$^2$.V$^{-1}$.s) &		500 &	800  &4500		 &		3000 \\
T$_F$(K) &		1150&	900  &87		 &		100 \\
$\frac{S^m}{T}$($\mu V$.K$^{-2}$) &		-0.4&	-0.45 &		 -6.1&-6.5 \\

$\frac{\nu^m}{T}$($\mu V$.K$^{-2}$.T$^{-1}$)  &		0.009&	0.012  &1.24		 &		2.4 \\
\hline
-$\frac{\pi^2}{2}\frac{k_B}{e}\frac{1}{T_F}$ &		-0.2&	-0.5  &-4.9		 &	-4.2	 \\
 283$\frac{\mu}{T}$($\mu V$.K$^{-2}$.T$^{-1}$)  &		0.009&	0.02  &1.44		 &		0.86 \
\end{tabular}
\end{ruledtabular}
\label{TabI}
\end{table}

\subsection{ Investigation of the quantum limit of the bulk  states of Bi$_2$Se$_3$ }

When the carrier concentration becomes as low as 10$^{17}$cm$^{-3}$, the so called \emph{quantum limit} can be attained by a magnetic field of 15 T. In this limit all carriers are confined to their lowest Landau level. One essential parameter in this regime is the magnitude of the $g$-factor. In the case of bismuth, for example,  when the field is oriented along the trigonal direction, the $g$-factor of the hole pocket is as large as 62. This corresponds to a ratio of the Zeeman energy (E$_Z$) to the cyclotron energy ($\hbar \omega_c$) labeled $M=\frac{E_Z}{\hbar \omega_c}$  slightly larger than 2 \cite{Bomparde2011,zhuPRB}. As a consequence, the chemical potential starts to move significantly with increasing magnetic field, dramatically affecting the Landau level spectrum \cite{{zhuPRB}} .

In the case of Bi$_2$Se$_3$, the data of sample B$_2$ down to T=0.38mK and up to B=17T does not reveal any Zeeman splitting of the peaks like the one found in bismuth \cite{Bomparde2011, behnia2}. This suggests that the Zeeman energy and the cyclotron energy are commensurate (i.e that M is closed to an integer in the limit of the peak width) as originally suggested by K\"{o}lher \emph{et al.}\cite{Kohler}. In order to differentiate between M=0,1,2,3,..., we used a model similar to the one used by K\"{o}lher \emph{et al.}\cite{Kohler} that determines the Landau level spectrum and the field dependence of the Fermi energy. The calculation of E$_F$ is performed with the assumption that the carrier concentration (noted $n$) is independent of the magnetic field (see Appendix \ref{HallPart} for a justification):  $n=\int^{E_F}_{-\infty}D(\epsilon)d\epsilon$ where $D(\epsilon)$ is the density of state, equal to the sum of the density of state of each Landau level (noted LL) depending on their position relative to the Fermi energy (see Appendix \ref{model} for more details). 
% alpha_{xy} or S_{yx) : 
\subsubsection{$ \alpha_{xy}$ or $N$ ? }

\begin{figure}[h]
\resizebox{!}{0.22\textwidth}{\includegraphics{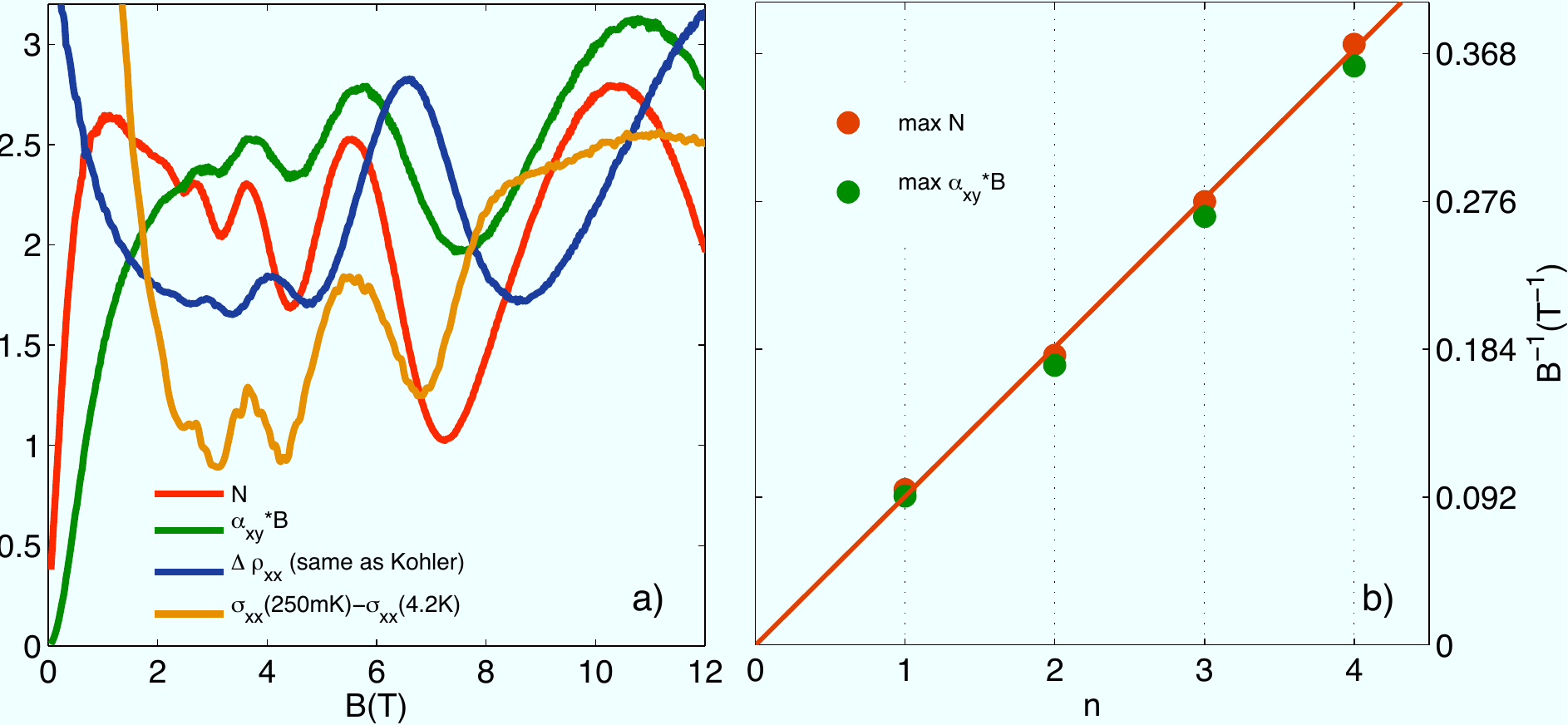}}\caption{ (Color online) a) Field dependence of  the Nernst effect at T=3K,  $\alpha_{xy}*B$(T=3K), $\Delta \rho_{xx}$  and $\Delta \sigma_{xx}$=$\sigma_{xx}(0.25K)-\sigma_{xx}(4.2K)$ respectively in red, green, blue and dark orange . b) Landau-level fan diagram for oscillations in N and $\alpha_{xy}*B$ respectively red and green. The red line corresponds to a linear fit of the red points. As seen in a), the peak positions in N  and $\alpha_{xy}*B$ are almost merging}
\label{Fig1SI}
\end{figure}

One complication arises in comparing the experimental Nernst peaks and the theoretical Landau spectrum. Several theories were proposed to explain the observation of giant quantum oscillations in bismuth and graphite  \cite{Bergman2010,Sharlai2011,Lukyanchuc2011}. Two of these theories focus on the off-diagonal thermoelectric conductivity $\alpha_{xy}$ \cite{Bergman2010,Sharlai2011}, another one is based on a semi classical description of the Nernst effect (N) \cite{Lukyanchuc2011}. In the case of bismuth and graphite, these two descriptions are equivalent because $\rho_{xx} >>\rho_{xy}$ and S$<<$N. However, in the case of Bi$_2$Se$_3$  $\rho_{xx}$ and $\rho_{xy}$ become comparable in amplitude and therefore $\alpha_{xy}\times B$ and $ N$ do not peak exactly at the same magnetic field. We report on Fig.\ref{Fig1SI} a) the field dependence of the four quantities: $\alpha_{xy}*B$, N, $\Delta \rho_{xx}$ and $\Delta \sigma_{xx}$=$\sigma_{xx}(0.25K)-\sigma_{xx}(4.2K)$ (where $\sigma_{xx}$=$\frac{\rho_{xx}}{\rho^2_{xx}+\rho^2_{xy}}$)  as a function of the magnetic field. Several conclusions can be drawn from this plot:

First, $\alpha_{xy}*B$ and the Nernst effect do not peak exactly at the same magnetic field. Fig.\ref{Fig1SI} b) shows the B$^{-1}$ position of the Landau levels  vs. the index number. As seen in Fig.\ref{Fig1SI} a), the difference between the peak positions in $\alpha_{xy}*B$ and in N is very small compare to the periodicity of the signals. Thus, our conclusions is not affected by the choice of $\alpha_{xy}*B$ or the Nernst effect. Further theoretical investigations are needed to clarify this issue.

Second, the maxima of $\Delta \sigma_{xx}$ are concomitant with the peak positions in $\alpha_{xy}$ and the Nernst effect, whereas the maxima in $\Delta \rho_{xx}$ are not. The origin of this difference is again related to the similar amplitudes of $\rho_{xx}$ and  $\rho_{xy}$ above a few Tesla. Thus it appears that the enhancement of the conductivity generated by the crossing of a LL (Landau level) and the Fermi energy leads to a maximum in the Nernst response in the case of Bi$_2$Se$_3$, as it is the case for graphite and bismuth.

In Fig.\ref{Fig1SI} b) (Landau level fan), the intercepts of $\alpha_{xy}*B$ and the Nernst effect are very close to 0, which is reminiscent of the non trivial Berry phase as observed in LaRhIn$_5$  \cite{Mikitik04}. In the case of Bi$_2$Se$_3$, which is characterized by a parabolic dispersion, the large spin orbit interaction generates this peculiar behavior. The occurrence of a vanishing intercept  points out to an odd value for the ratio of the Zeeman energy and the cyclotronic energy (labelled M). This value is different from that deduced from the analysis of the peak position of $\Delta \rho_{xx}$ by K\"ohler et al. \cite{Kohler} (M=2). However, as discussed previously, it has been shown that the peak position in $\Delta \rho_{xx}$ differs from the Nernst response (and also differs from $\sigma_{xx}$). In addition, as we approach the quantum limit, the spectrum analysis becomes more difficult because of the field dependence of the Fermi energy. In order to clarify the possible values of M we proposed in the next section a detail analysis of the Nernst peaks using the simple model introduce in  Appendix \ref{model}).

% Facteur g : 

\subsubsection{ $g$-factor of the bulk states of Bi$_2$Se$_3$}

\begin{figure}[h]
\resizebox{!}{0.45\textwidth}{\includegraphics{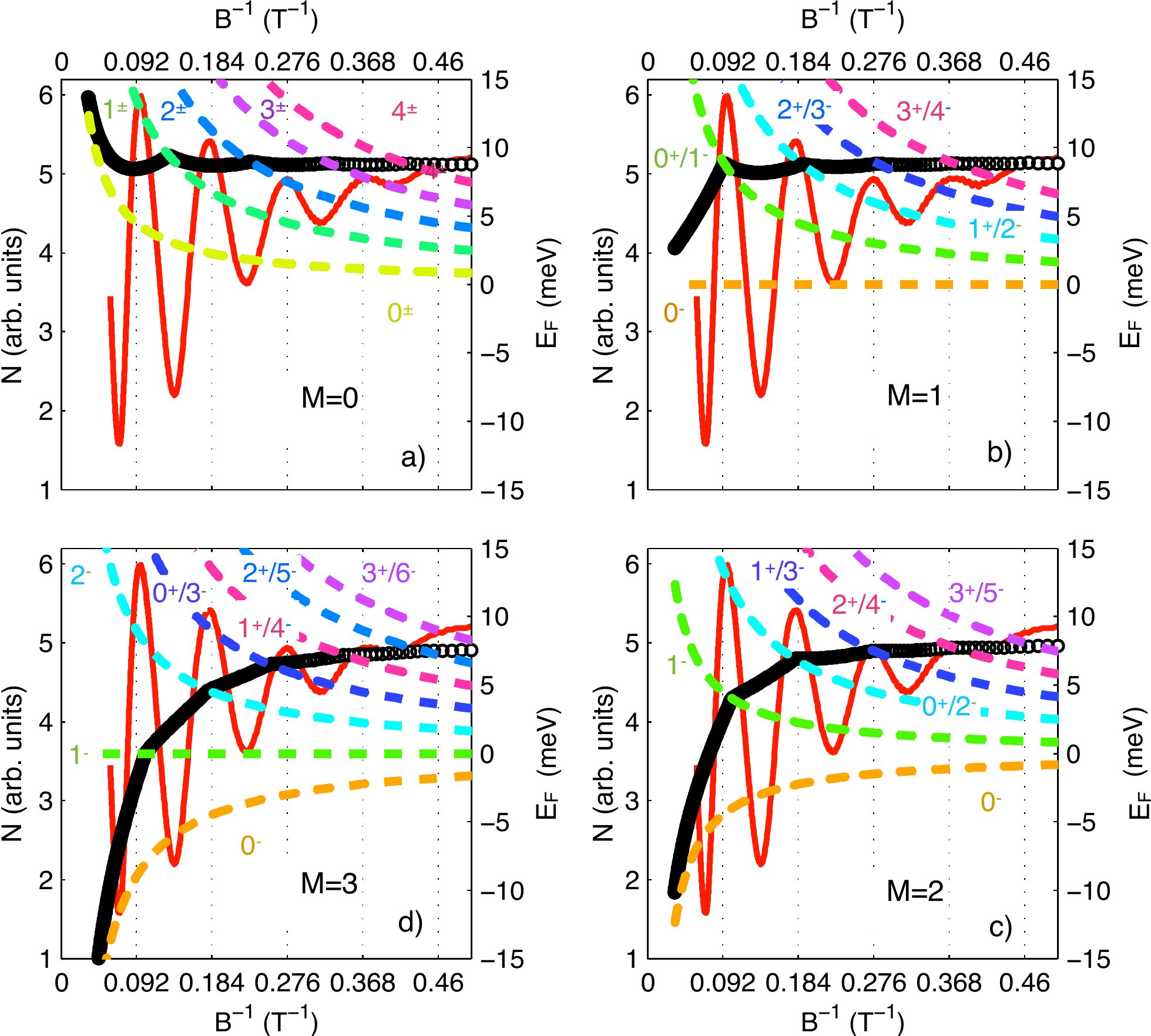}}\caption{ (Color online) Nernst effect (red line) of the sample B$_1$ at T=3K  on top of a calculated Landau level energies and Fermi energy E$_F$ (black open circle) with different parameters: M=0,1,2 and 3 which respectively correspond to a), b), c) and d).  The calculation was performed with a carrier concentration of n=1.08 $10^{17}cm^{-3}$ for a), b) and d)  and n=1.25 $10^{17}cm^{-3}$ for b)  , a Dingle temperature of T$_D$=8K and m$_1$=m$_2$=0.14m$_0$ and m$_3$=0.24m$_0$.  }
\label{Fig2SI}
\end{figure}

The last peak resolved in the Nernst effect in sample B$_1$ occurs at 10.4T. We have adjusted the carrier concentration in order to find the best agreement between the position of this peak and the crossing points between a Landau level and the Fermi energy for different possible values of M. Fig.\ref{Fig2SI} a), b), c) and d) shows the Nernst effect (red lines) as function of B$^{-1}$, superimposed with calculated Landau levels and Fermi energy E$_F$ (black points) with various parameters: M=0,1,2 and 3.

% cas M=0 : 

\paragraph{M=0 : }

\begin{figure}[h]
\resizebox{!}{0.215\textwidth}{\includegraphics{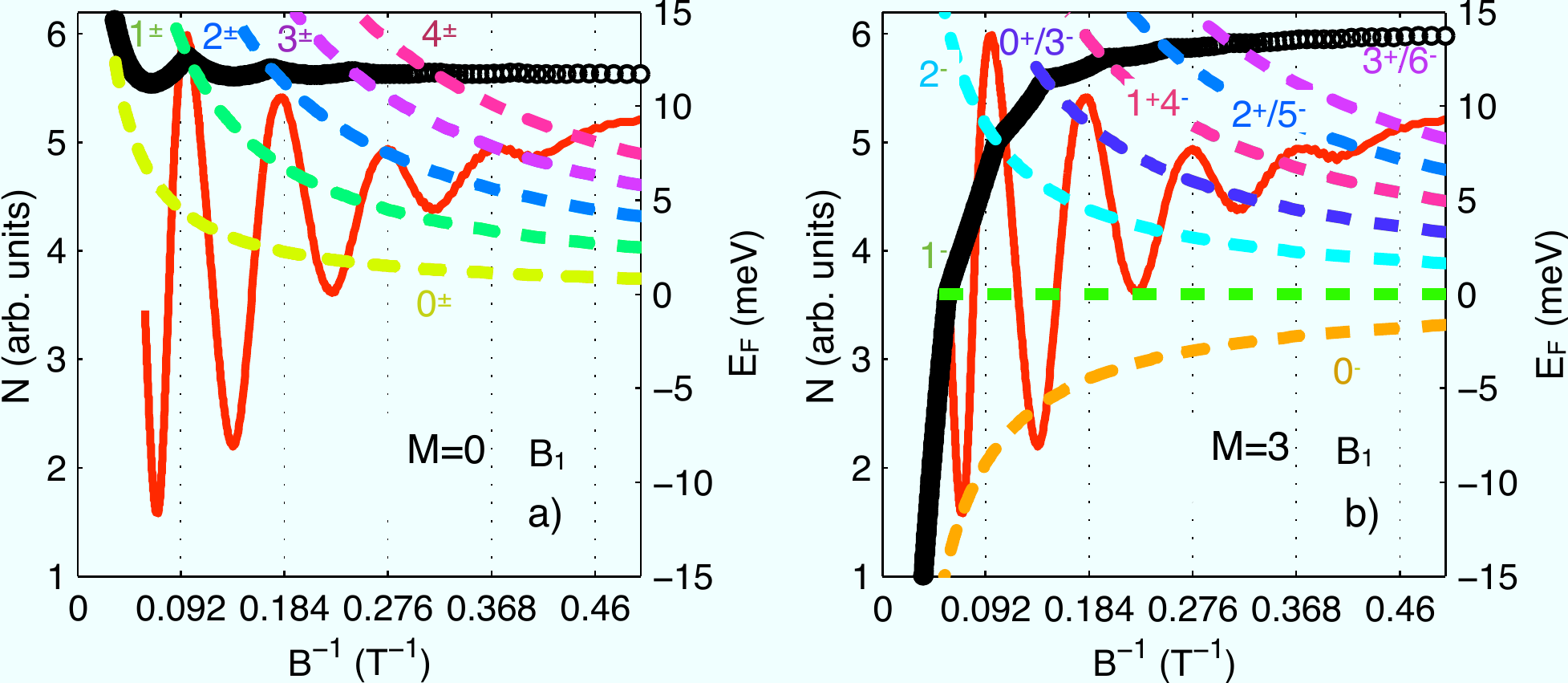}}\caption{  (Color online) N (in red lines) of the sample B$_1$ at T=3K on top of a calculated Landau level energies and Fermi energy E$_F$ (black open circle) with M=0 for a) and M=3 for b). In a case of  a) the carrier concentration has been chosen in order that the last resolved peak corresponds to the LL 1$^{\pm}$ ($n$=1.8 $10^{17}cm^3$).  For b), the carrier concentration has been chosen in order that the last resolved peak matches the LL 2$^-$ ($n$=2.9 $10^{17}cm^3$). The calculation was performed with a Dingle temperature T$_D$=8K and m$_1$=m$_2$=0.14m$_0$ and m$_3$=0.24m$_0$ }
\label{Fig3SI}
\end{figure}

On Fig.\ref{Fig2SI} a), we adjust the carrier concentration to match the periodicity of the Nernst response ($n$=1.08  $10^{17}cm^{-3}$). In this case, the crossings of the LL and the Fermi energy correspond to minima and not maxima. We cannot reproduce the lowest minimum of the Nernst effect. In Fig.\ref{Fig3SI} a), we try to adjust the carrier concentration to match the last resolved peaks at 10.4T and the LL 1$^{\pm}$. However, with this carrier concentration ($n$=1.8 $10^{17}cm^{-3}$), the low-field peaks cannot be reproduced at the right positions.

\paragraph{M=1 : }
Fig.\ref{Fig2SI} b) shows the Landau level spectrum for a bulk carrier concentration of $n$=1.28 $10^{17}cm^{-3}$. In this case, there is not a complete agreement for the two last peaks, but it is rather good for the other peaks. In the case of M=1, the LL n+1$^{-}$ and n$^{+}$ are degenerate except for the last ones:  0$^-$ (in orange on Fig.\ref{Fig4} a)). The last observed peak at 10.4T is attributed to the Landau levels 0$^{+}$/1$^{-}$. Above this field all the electrons are in the lowest Landau level   0$^-$.

\paragraph{M=2 : }
Fig.\ref{Fig2SI} c) shows the Landau level spectrum for a bulk carrier concentration of $n$=1.08 $10^{17}cm^{-3}$. We find a rather good agreement for all the peaks. Note that the agreement is slightly better for M=2 than for M=1. In the case of M=2,  the LL n+2$^{-}$ and n$^{+}$ are degenerate except for the two last ones 1$^-$ (in green on Fig.\ref{Fig4} b)) and 0$^-$ (in orange on Fig.\ref{Fig4} b)). The last observed peak at 10.4T is then attributed to the Landau level 1$^{-}$ and the Landau levels n$^{-}$ are degenerate with n-2$^{-}$. This result is compatible with the conclusion of the early Shubnikov-de Haas analysis \cite{Kohler}. 

\paragraph{M=3 (and above) : }
We also investigated higher values for M. On Fig.\ref{Fig2SI} d) (M=3), we report the Landau level spectrum for a carrier concentration of n=1.08 $10^{17}cm^{-3}$. The lowest observed peaks match well the 1$^{-}$ LL. This result is rather surprising as we would have naively expected that the field scale associated with the 1$^{-}$ LL would continuously increase with the value of M. In fact, even if  the two lowest LL 1$^{-}$, 0$^{-}$ and the chemical potential are decreasing faster for M=3 than for M=2, the crossing point between the  LL 1$^-$ and the chemical potential occurs at the same field for M=2 and M=3. This conclusion is true for all the values M$>2$.  The situation is however different for the other LL, where the crossing points differ for M=2 and for M=3. In other words, for M=3 we can find a carrier concentration where the 10.4T peak matches the LL 1$^-$, however the peak positions at lower field cannot be explained. On Fig.\ref{Fig3SI} b), we investigate the possibility that the lowest observable peak is not associated with the 1$^-$ LL but with the 2$^-$ LL. We found a carrier concentration of $n$=3 $10^{17}cm^3$. However, the peak positions at lower fields cannot be explained.

In conclusion, the M=1 and M=2 hypothesis yields the best agreement  between the peak positions in the Nernst effect and the crossing of Landau levels and the Fermi energy. Remarkably, in both cases above 10.4T,  all bulk carriers are spin polarized. For M=1 and M=2 and with a cyclotronic mass of 0.14m$_0$, we find respectively that g=14.3 and g=28.6.\\

\subsubsection{ Nernst effect at high field}

Sample B$_{3}$ was  studied up to 32T. In this case, the last resolved peak in the Nernst effect occurs at 11.4T because of a slightly higher carrier concentration. Interestingly, between 17T to 32T, an unexpected increase of the Nernst effect  was observed. At first glance, one could explain this increase by the presence of additional Landau levels which are expected in the case of M$>$2. For instance for M=3, the last observed peak could be attributed to the LL 2$^-$. However, this scenario does not reproduce the low field peak positions. Moreover, it would yield a chemical potential crossing for the 1$^-$  LL  at 19T, in contrast with our experimental data. 

\begin{figure}
\resizebox{!}{0.215\textwidth}{\includegraphics{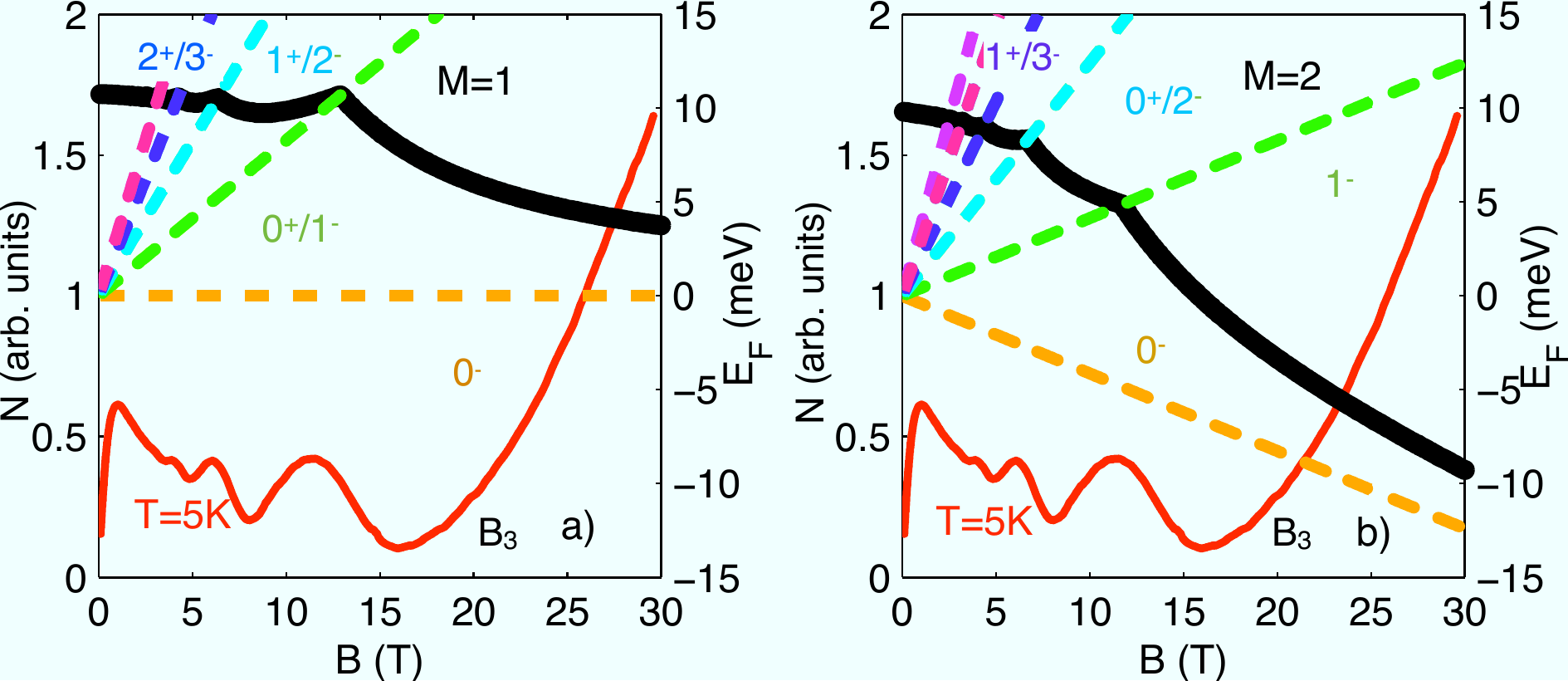}}\caption{ (Color online) In red line is the Nernst effect, $N$, as a function of B$^{-1}$ at T=3K for the sample B$_3$( same as B$_1$ few months lalter) on a) and b). The evolution of the Fermi energy (E$_F$) is plotted in black points. Each time that the chemical potential is crossing a Landau level (colors lines), there is a kink due to the carrier variation. The best parameter to describe the peaks position as measured  in $N$ is found to be for M=1 (with a  bulk carrier concentration n=1.8 10$^{17}$cm$^{-3}$)  on a) and M=2 ( with a bulk carrier concentration n=1.5 10$^{17}$cm$^{-3}$) on b) .}
\label{Fig4}
\end{figure}

The increase in the Nernst effect observed at high field calls for an alternative scenario. As reported on Fig.\ref{Fig4} b), once the  1$^-$ LL crosses the chemical potential,  the enhanced shift in chemical potential pushes it closer towards the 0$^{-}$ LL. In the the low field regime the Nernst response increases as the distance between the highest filled Landau level and the chemical potential becomes shorter and attains a maximum  when the two levels coincide. Here if  the chemical potential approaches the 0$^{-}$ LL without crossing it in order to keep charge neutrality, one expects it to peak without attaining a peak. This scenario can qualitatively explain the experimentally-observed increase in the Nernst effect.\\

\subsection{ Conclusion  }

In conclusion,  the overall thermoelectric and thermomagnetic response in Bi$_2$Se$_3$ can be quantitatively understood in the context of a single 3D band in presence of large Zeeman splitting. Our investigation reveals a large variation of  thermoelectric and thermomagnetic coefficients with the bulk carrier concentration which can be quantitatively understood as a result of a change in the mobility and the bulk Fermi energy. We resolved large quantum oscillations in the Nernst coefficient suggesting that such an effect is a common feature to both compensated and non compensated low carrier systems. For the lowest carrier concentrations investigated, we found that the Zeeman energy is either equal to one or two  times the cyclotron energy. In both case, a strong variation of the chemical potential above the quantum limit is expected, which can naturally explain the observed increase in the Nernst response. 
 \color{black}
Finally we would like to point out that our results could impact the physics of the surface states. In the presence of a metallic bulk coupled with surface states, we can expect that the field-induced shift of the chemical potential of the bulk state could also generate a variation of the Fermi energy of the surface states. Such a situation will affect, for example, the Landau level indexing and the determination of the Berry phase of the surface state
\color{black}

This work is supported by the Agence Nationale de Recherche as a part of the QUANTHERM project, by EuroMagNET II under the EU contract no. 228043. Work at Maryland was partially supported by the Lawrence Livermore National Laboratory LDRD program (Tracking Code 11-LW-003), operated under DOE-NNSA Contract No. DE-AC52-07NA27344.

 %\appendix

% \begin{widetext}
 
 \appendix

\appendix

\section{Hall effect and carrier concentration}
\label{HallPart}

%\subsection{Sample growth}
%Single crystals of Bi$_2$Se$_3$ were prepared by melting high purity bismuth (6N) and selenium (5N) in sealed quartz ampoules as previously reported [S1].  The carrier density %was controlled by varying the initial Bi:Se ratio.

%\subsection{Hall effect and carrier concentration}

\begin{figure}[h]
\resizebox{!}{0.22\textwidth}{\includegraphics{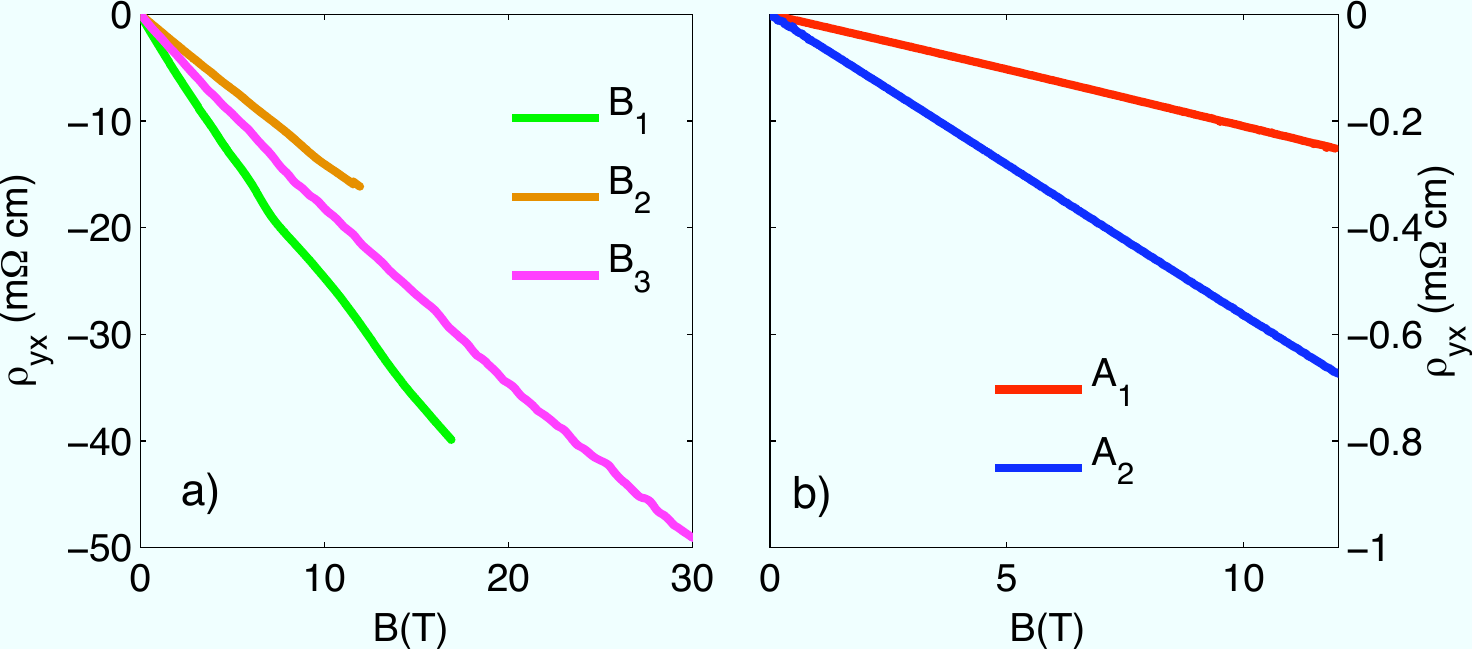}}\caption{ a) Field dependance $\rho_{yx}$ for B$_1$ (T=1.5K up to 12T), B$_2$ (T=0.35K up to 17T)  and B$_{3}$ (T=1.6K up to 30T) b) Field dependance $\rho_{yx}$ for A$_1$ and A$_2$ at T=1.6K up to 12T }
\label{FigSIHall}
\end{figure}

 Fig.\ref{FigSIHall} shows the field dependence of the Hall resistivity for five samples. Up to 32T and down to the lowest carrier concentration investigated, the Hall effect was found to be linear in magnetic field suggesting that the carrier concentration remains constant and does not vary with the field. From the slope, we can determine the carrier concentration, labeled n$^H$. These values can be directly compared with the carrier concentration deduced from the period of the quantum oscillations labeled n$^F$. The conduction-band structure in this range of carrier densities is approximately parabolic.  The Fermi energy is then determined by the period of the oscillation through the Onsager relation. Assuming a mass anisotropy independent of the doping ( $m_1$=$m_2$=0.18m$_0$ and m$_3$=1.7$m_1$) one can determine  the carrier concentration of the samples. For all samples, within few pourcent, we found a good agreement between n$^F$ and n$^H$.
%. the carrier concentrations determined by the two methods are in good agreement.
% \begin{widetext}
 %\begin{table}[h]
 %\caption{\label{tab:table1} Carrier concentration and electronic mobility deduced from various transport measurement. n$^H$ and n$^F$ correspond respectively to the carrier %concentration deduced from the Hall effect and from quantum oscillations of the Nernst effect.}
 %\begin{ruledtabular}
%\begin{tabular}{lclccccl}
%Sample &A$_1$\footnote{measured down to 2K}&A$_2$\footnote{measured down to 0.15K}&B$_{1}$$^b$&B$_2$$^a$ &B$_{3}$$^a$\\
%\hline
%n$^H$(cm$^{-3}$) &	$3\times 10^{19}$	 &	$1\times 10^{19}$  &$	2.5\times 10^{17}$	 &	$4\times 10^{17}$ &$3\times 10^{17}$	 \\
%n$^F$(cm$^{-3}$) &		$2.4\times 10^{19}$ &$1.5\times 10^{19}$  &$3.8\times 10^{17}$		 &$4.8\times 10^{17}$ &	$4.4\times 10^{17}$ \\
%\hline
%$\mu(Magnetoresistance)$ (cm$^2$.V$^{-1}$.s)&	500	 &	800  &	4500	 &	5000 &$\approx$4500 \\
%$\mu(Hall)$ (cm$^2$.V$^{-1}$.s)& 670		&	1200 	&	4740 	& 5330		&	$\approx$4000	 \\
%\hline
%\end{tabular}
%\end{ruledtabular}
%\label{SITab1}
%\end{table}
 %\end{widetext}
\section{ Electronic mean free path and impurity scattering}
%\label{HallPart}
%\subsection{Electronic mean free path and impurity scattering}

In the case of Bi$_2$Se$_3$ it is believed that Se vacancies are the charge dopants responsible for the n-type nature of the system. Here we propose to make a simple discussion to know if the vacancies are the main source of scattering of the electrons at low temperature. In other words we propose to compare the electronic mean free path and the distance between vacancies (d$_v$). For the two extreme concentration that we studied, we can determine the electronic mean free path $l_e$ using the knowledge of the mobility and the Fermi velocity : $l_e$=v$_F$$\tau$=$\frac{\hbar k_F\mu}{e}$.

Each Se atoms can contribute for two electrons. For a given bulk carrier concentration $n$, the number of lacuna per unit cell, $N_i$ is given by : $N_i=\frac{nV}{2}$ where  V the unit volume. If we assume a homogeneous distribution of the lacuna, the typical distance between vacancies sites can be estimated  : $\frac{4}{3}\pi d_v^3$= $\frac{V}{N_i}=\frac{1}{n}$. Tab.\ref{SITab2} yields values of d$_v$ for the two carrier concentrations. For both samples, the mean-free-path is longer than the typical distance between impurity sites, suggesting that the scattering probability between a traveling electron and a point defect is much smaller than unity.
 
 \begin{table}[h]
 \caption{\label{tab:table1} Comparison of the electronic mean free path and the distance between vacancies sites}
 \begin{ruledtabular}
\begin{tabular}{lclccccl}
Sample & n (cm$^{-3}$)& l$_e$(nm) & d$_v$ (nm)\\
\hline
A$_1$ &  $3\times 10^{19}$& 23 & 2	 \\
B$_1$ & $2.5\times 10^{17}$ &550 & 10	 \\
\end{tabular}
\end{ruledtabular}
\label{SITab2}
\end{table}

%\subsection{Dingle Temperature}

%\subsection{$\alpha_{xy} \times B$ or N ?}

\section{ Field dependence of the Fermi energy }
\label{model}

 In presence of a large $g$-factor (M$\geqslant$1) for one single band, the Landau spectrum is modified in two ways:
 
 i) The quantum limit is reached at a higher magnetic field. For example when M=2, the quantum limit is approximatively one period above the quantum limit attained than in the case of M=0.  
 
 ii)  The lowest LL 0${^-}$ is going down with increasing magnetic field. This pulls down the chemical potential near the quantum limit regime. In order to quantify these effects in the case of Bi$_2$Se$_3$, we compute the field dependence of  the Fermi energy for various value of M.  As discussed in the Appendix \ref{HallPart}, we assumed that the carrier concentration is independent of the magnetic field and is given by  $n=\int^{E_F}_{-\infty}D(\epsilon)d\epsilon$ where $D(\epsilon)$ is the density of state.  In presence of a magnetic field and in the absence of a spin splitting, the density of state $D_0(\epsilon)$ can  be written as :

 \begin{widetext}
 \begin{equation}
	D_0(\varepsilon) =
		\frac{\sqrt{m_{0}}}{2 \pi^{2} \hbar^{2}} \hspace{3pt} eB
		\sum_{n=0}^{n_{max}}	\left [ \frac{ \varepsilon - \varepsilon_{n} +
						\sqrt{ \vphantom{\frac{1}{2}}(\varepsilon - \varepsilon_{n})^{2} + \Gamma^{2}} }
						{ (\varepsilon - \varepsilon_{n})^{2} + \Gamma^{2}}  \right ]^{1/2} \times \hspace{5pt}
	\begin{cases}
		1
		&\text{ if } \varepsilon > \hbar \omega_{c} /2  \\
 		0
		&\text{ if } \varepsilon < \hbar \omega_{c} /2
	\end{cases}
	\label{eDOSDingle}
\end{equation}
\end{widetext}

Where $\epsilon_n$=$(n+\frac{1}{2})\hbar \omega_c$, $n_{max}$ is the highest value of $n$ yielding a positive value value for $\varepsilon - \varepsilon_{n}$ and $\Gamma$ is the broadening of the Landau levels ($\Gamma$=$\pi \frac{k_B T_D}{e}$). In presence of a spin splitting $\epsilon_n$=$(n+\frac{1}{2}\pm\frac{M}{2})\hbar\omega_c$ where $M=\frac{gm_c}{2m_0}$. In this case, the density of state is given by :  $D(\epsilon)=\frac{1}{2}(D_0(\epsilon-\frac{M}{2}\hbar \omega_c)+D_0(\epsilon+\frac{M}{2}\hbar \omega_c))$.

%\emph{Reference Supplementary Information}  

%[S1] N. Butch et al., Phys. Rev. B ${\bf{81}}$, 241301(R) (2010)

%[S2] D. Shoenberg, Magnetic Oscillations in Metals, Cambridge University Press, Cambridge (1984)

%[S3]  H. K\"ohler and E.W\"uchner, Phys. Stat. Sol. ${\bf{67}}$,665 (1975)

%[S4] K. Eto et al., Phys. Rev. B ${\bf{81}}$, 195309 (2010)

%[S5]  J. G. Analytis et al.,  Phys. Rev. B ${\bf{81}}$, 205407 (2010)

%[S6] D. L. Bergman and V. Oganesyan , Phys. Rev. Lett. ${\bf{104}}$, 066601 (2010)

%[S7] Y. V. Sharlai and G. P. Mikitik Phys. Rev. B ${\bf{83}}$, 085103 (2011)

%[S8] I. A. Lukyanchuk et al., Phys. Rev. Lett. ${\bf{107}}$, 016601 (2011)

%[S9] G.P Mikitik and Yu. V Sharlai,  Phys. Rev. Lett. ${\bf{93}}$, 106403 (2004)
%\end{widetext}


\begin{thebibliography}{99}

\bibitem{Hasan} M.Z.Hasan and C.L Kane, Rev. Mod. Phys. {\bf{82}}, 3045 (2010)
\bibitem{NolasBook} G.S Nolas, J. Sharp, H.J Goldsmid, Thermoelectrics, Springer, (1962)
\bibitem{Goldsmith} H.J. Goldsmid, Proc. Phys. Soc. London {\bf{71}}, 633 (1958)
\bibitem{Mahan} G. D. Mahan, Solid State Phys. \textbf{51}, 81 (1998)
%\bibitem{Zhang} H.Zhang et al., Nature Physics {\bf{5}}, 438 (2009)
\bibitem{Butch} N. Butch et al, Phys. Rev. B {\bf{81}}, 241301(R) (2010)
\bibitem{Kohler} H. K\"ohler and E.W\"uchner, Phys. Stat. Sol. ${\bf{67}}$, 665 (1975)
\bibitem{Ando} Kazuma Eto, Zhi Ren, A. A. Taskin, Kouji Segawa, and Yoichi Ando, Phys. Rev. B {\bf{81}}, 195309 (2010)
\bibitem{Analtyis} James G. Analytis et al.,  Phys. Rev. B {\bf{81}}, 205407 (2010) 

\bibitem{behnia2} K. Behnia, M. -A. M\'easson and Y. Kopelevich, Phys. Rev. Lett. \textbf{98}, 166602 (2007)
\bibitem{zhu1}   Zengwei Zhu, Huan Yang, Beno"t FauquŽ, Yakov Kopelevich and Kamran Behnia, Nature Phys., \textbf{6} 26 (2010)



\bibitem{Bergman2010} D. L. Bergman and V. Oganesyan , Phys. Rev. Lett. ${\bf{104}}$, 066601 (2010)
\bibitem{Sharlai2011} Y. V. Sharlai and G. P. Mikitik Phys. Rev. B ${\bf{83}}$, 085103 (2011)
\bibitem{Lukyanchuc2011} Igor A. LukÕyanchuk, Andrei A. Varlamov, and Alexey V. Kavokin, Phys. Rev. Lett. ${\bf{107}}$, 016601 (2011)

\bibitem{Bergeron59} C. J. Bergeron, C. G. Grenier, and J. M. Reynolds, Phys. Rev. Lett. {\bf{2}}, 40 (1959)
\bibitem{Fletcher83} R. Fletcher, Phys. Rev. B {\bf{28}}, 6670 (1983)
\bibitem{Gadzhialiev69} M.M Gadzhialiev, Sov. Phys. Semicon. {\bf{2}}, 833 (1969)
\bibitem{Tieke96}  B. Tieke, R. Fletcher, J. C. Maan, W. Dobrowolski, A. Mycielski, and A. Wittlin, Phys. Rev. B {\bf{54}}, 10565 (1996)
\bibitem{Shoe} D. Shoenberg, Magnetic Oscillations in Metals, Cambridge University Press, Cambridge (1984)
\bibitem{BehniaNernst} K. Behnia, J. Phys. Condens. Matter {\bf{21}}, 113101 (2009)
\bibitem{BehniaSeebeck} K. Behnia, D. Jaccard, J. Flouquet, J. Phys.:Condens. Matter {\bf{16}}, 5187(2004)
\bibitem{Bomparde2011} S. G. Bompadre, C. Biagini, D. Maslov, and A. F. Hebard , Phys. Rev. B {\bf{64}}, 073103 (2001)
\bibitem{zhuPRB} Zengwei Zhu, Beno"t Fauqu\'e, Yuki Fuseya, and Kamran Behnia , Phys. Rev. B {\bf{84}}, 115137 (2011)
 \bibitem{Mikitik04} G.P Mikitik and Yu. V Sharlai,  Phys. Rev. Lett. ${\bf{93}}$, 106403 (2004)
 \end{thebibliography}
\end{document}